**AI Sprints: Towards a Critical Method for Human-AI Collaboration**

David M. Berry

University of Sussex

**Abstract**

The emergence of Large Language Models presents a remarkable opportunity for humanities and social science research. I argue these technologies instantiate what I have called the "algorithmic condition", whereby computational systems increasingly mediate not just our analytical tools but how we understand nature and society more generally. This article introduces the possibility for new forms of humanistic inquiry through what I term "AI sprints", intensive time-boxed research sessions. This is a research method combining the critical reflexivity essential to humanistic inquiry with iterative dialogue with generative AI. Drawing on experimental work in critical code studies, I demonstrate how tight loops of iterative development can adapt data and book sprint methodologies whilst acknowledging the profound transformations generative AI introduces. Through examining the process of human-AI collaboration when undertaken in these intensive research sessions, I seek to outline this approach as a broader research method. The article builds on Rogers' digital methods approach, proposing that we extend methodologies to study digital objects through their native protocols, using AI systems not merely to process digital traces but to analyse materials traditionally requiring manual coding or transcription. I aim to show this by introducing three cognitive modes, cognitive delegation, productive augmentation, and cognitive overhead, explaining how researchers can maintain a strategic overview whilst using LLM capabilities. The paper contributes both a practical methodology for intensive AI-augmented research and a theoretical framework for understanding the epistemological transformations of this hybrid method. A critical methodology must therefore operate in both technical and theoretical registers, sustaining a rigorous ethical-computational engagement with AI systems and outputs.

**Introduction**

The emergence of Large Language Models (LLMs) presents a remarkable opportunity for humanities and social science research. As I have argued elsewhere, we face an "algorithmic condition" where computational systems increasingly mediate not just our analytical tools but how we understand culture and society more generally (Berry 2025). This also opens the possibility for new methodological approaches capable of exploiting AI's augmentative capacities whilst maintaining the critical reflexivity that is key to humanistic inquiry.

The method I propose responds to what might be understood as an augmentative moment in computational research. I argue that LLMs represent new tools through what Agre might call a shift in the "grammar of action" through which computational systems model domains of practice (Agre 2008).[1] Where earlier digital methods required researchers to translate analytical questions into procedural operations through coding or configuration of a digital tool, LLMs enable something closer to what we might call *conversational computation*, where natural language prompts mediate between research intentions and computational systems. This shift carries epistemological implications that require new methods capable of using these capabilities whilst maintaining a human-in-the-loop of research activities.

The intervention proposed here responds to what can become an interpretive bottleneck in computational social science and digital humanities. Whilst digital methods have proliferated tools for data collection and processing, the manual labour of intermediate interpretation remains difficult to automate. LLMs are interesting to experiment with in research because that may offer relief from this bottleneck through their capacity to generate analytical framings, thematic categories, and useful intermediate interpretive support. Yet care must be taken when using LLMs as they may collapse the very distance between data and interpretation that makes humanistic inquiry critical rather than merely descriptive. The AI sprint approach I describe in this paper, treats LLMs as an augmentative infrastructure, maintaining human judgement whilst using computational capabilities for processing and generating intermediate objects and data.

In this paper I want to suggest that "AI sprints" might be a possible method for using those capacities, building upon my recent experiments in critical code studies where I have developed an educational application through iterative dialogue with generative AI (Berry 2025b). That experiment, which adapts the idea of "vibe coding" following Karpathy's (2025) neologism, revealed the suggestive potential of human-AI collaboration when undertaken in tight loops of iterative development. Here I want to outline this approach as a broader research method, one that adapts the data and book sprint idea whilst acknowledging the profound transformations generative AI introduces (see Berry et al 2015).

---

[1] Agre's concept of a grammar of action describes how computational systems are designed to impose structures on the domains they model, creating a duality which includes the reciprocal relationship between computational representations and the practices they try to capture.

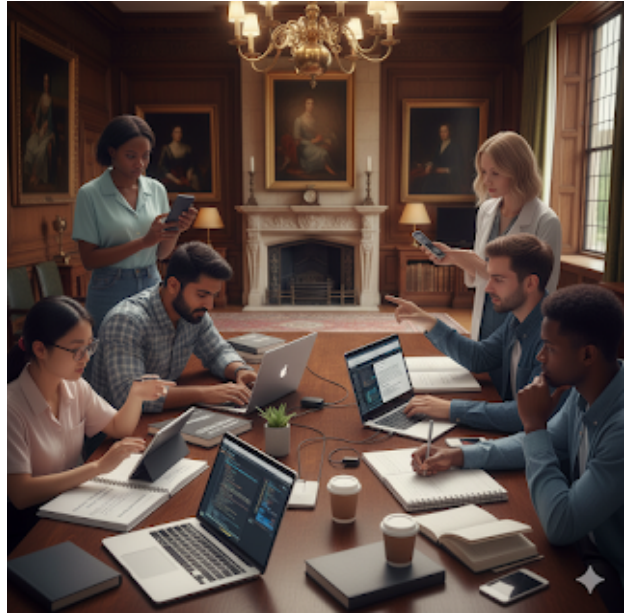

*Figure 1: Gemini generated image[2]*

As Rogers (2013) argues, digital methods allow us to "follow the medium", studying digital objects through their native protocols and using the strengths of the digital in order to develop research projects and methods. The *AI methods* I propose here extend this principle, using computational systems not merely to process digital traces but to analyse materials that traditionally required extensive manual coding or transcription. This could include AI-supported qualitative coding of interview transcripts, content analysis of archival materials, thematic analysis across large text corpora, and extraction of structured data from semi-structured documents. However, this extension risks "cognitive delegation", which I describe as the offloading of interpretive judgement to systems that lack human understanding. A critical methodology must therefore operate in both technical and reflexive registers, utilising AI's strengths whilst critiquing its epistemological assumptions and outputs.

---

[2] image generated using Google Gemini Pro. November 2025. The prompt used was: "draw a high-resolution, photographic image capturing the concept of an 'AI Sprint.' The scene is set in a classic Oxford senior common room, characterised by dark wood paneling, floor-to-ceiling bookshelves with leather-bound books, and large, leaded glass windows. Around a large, polished wooden table, a diverse group of six students and researchers are deeply engaged in collaborative work. They are using a mix of modern and traditional research tools: laptops, smartphones, paper notebooks, and books are all visible and in use. The group is in discussion, pointing at screens and notes, embodying an atmosphere of intense, focused, and critical inquiry. There is no large central presentation screen or monitor; the focus is entirely on the group's interaction with each other and their personal research materials." Due to the probabilistic way in which these images are generated, future images generated using this prompt are unlikely to be the same as this version.

**From Data Sprints to AI Sprints**

The "data sprint" methodology emerged from the notion of book sprints which were an agile practice for the rapid writing of bookish objects (see Berry et al 2015). In digital methods research, this has been taken further through the Digital Methods Initiative (DMI) at the University of Amsterdam. Traditional data sprints typically involve teams with mixed technical and domain expertise, intensive time-boxed work creating productive constraints, iterative cycles between data collection, processing, visualisation, and interpretation, production of "intermediate objects" such as graphs, networks, and timelines that mediate between raw data and analytical claims, and public presentation of provisional findings (Rogers 2013).

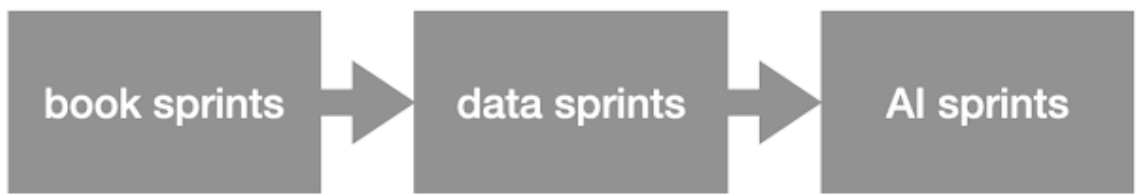

*Figure 2: the development of sprint forms: book sprint (2008), data sprints (2014) and AI sprints (2025)[3]*

The DMI data sprint approach relies particularly on what Rogers calls "issue mapping" and "controversy analysis", tracing how matters of concern circulate through digital media. These sprints have the potential to democratise digital research by creating structured environments where non-technical researchers can engage computationally-mediated inquiry through collaboration with technical partners. As Marres (2015) observes, this collaborative division of labour can be useful in digital sociology by distributing expertise across teams.

This potential shift in collaborative expertise towards individual researchers working with LLMs raises important questions about the social epistemology of digital research. Whilst AI augmentation may democratise access to computational methods by reducing dependence on technical collaborators, it might also threaten the productive friction that emerges when researchers with different epistemic commitments must negotiate shared analytical approaches or methods. The data sprint's strength lay not simply in distributing tasks but in requiring continual dialogue between differently positioned knowers. AI sprints risk losing this dialogical dimension, replacing it with what might be called a monological augmentation where a single researcher's interpretive approach remains unchallenged by alterity. This suggests AI sprints might work best when combined with traditional collaborative practices rather than replacing them entirely, using LLM augmentation to extend individual analytical capacity whilst maintaining spaces for collective interpretive work.

---

[3] Figure 2 shows the development of sprint forms across three distinct phases. Book sprints (2008) established intensive collaborative writing practices emerging from FLOSS documentation communities. Data sprints (2014) adapted this approach to digital methods research, introducing computational processing and visualisation within time-boxed collaborative work. AI sprints (2025) represent a further step, where LLM augmentation facilitates researchers working in tight iterative loops with generative AI systems. The diagram reveals both continuities in sprint methodology, such as intensive time constraints, iterative refinement, and the production of intermediate objects, and the potential reconfiguration of human-computer collaboration using LLMs.

AI sprints thus represent both a continuation and a break with the data sprint approach. We might also add that AI sprints respond to what might become a potential "methods crisis" in digital humanities and social sciences. This crisis emerges as LLMs become increasingly capable of performing many tasks that previously required bespoke computational tools, potentially undermining the rationale for developing field-specific digital methods expertise. If an LLM can generate code to analyse Twitter data, map controversies, or extract themes from interviews without requiring specialised scraping tools, network analysis software, or coding schemes, what is likely to become of the methodological fields built around such tools? AI sprints address this crisis not by abandoning digital methods but by integrating LLM capabilities within established methodological approaches that maintain disciplinary traditions that privilege interpretation and critique. The continuity lies in AI sprints emphasis on time-boxed intensive work, iterative refinement through feedback loops, production of intermediate mediating objects, and commitment to making research processes open and teachable. The difference emerges through the changed nature of human-computer collaboration. Data sprints distribute expertise across human teams, using technical researchers to handle scraping and processing, and domain experts guide interpretation, creating boundary objects that coordinate activity across different communities of practice (Star and Griesemer 1989).[4] In contrast, AI sprints potentially collapse this distribution towards individual researchers working in tight loops of dialogue with LLMs that handle each of these previous specialisms.[5]

This is not simply automation but what I term "cognitive augmentation", where the LLM extends research capacity whilst the human researcher maintains top-level control over the inquiry (Berry 2025b). As I discovered through developing my own software application using Google Gemini Pro, I found that this augmentation operates through three distinct modes which I call: (1) *cognitive delegation*, (2) *productive augmentation*, and (3) *cognitive overhead*. These cognitive modes are important for understanding the best ways of approaching an AI sprint method and determining when collaboration can be successful.[6]

---

[4] The concept of boundary objects describes artefacts that are plastic enough to adapt to local needs whilst maintaining coherence across sites of practice.

[5] The emergence of frontier LLMs poses what might become an existential challenge to the digital humanities. Much of the field has justified itself through appeals to a "methodological commons" of shared digital tools, techniques, and archives that require specialised technical literacy to access and deploy. If LLMs can generate bespoke analytical tools from natural language prompts and process archival materials without requiring such specialised knowledge, what becomes of these justifications for DH? The DH field is slowly becoming aware of this potential crisis, whilst continuing to work on tool development and infrastructure building without necessarily considering how generative AI might undermine the rationale for this activity. This reinforces the importance of a critical digital humanities capable not merely of applying digital methods but of continually interrogating their epistemological foundations and their embeddedness within computational capitalism. A critical approach pays close attention to how these AI methods emerge from and reinforce particular formations of power and capital, rather than treating computational tools as neutral instruments for humanistic inquiry. This requires engagement with how AI systems emerge from and reinforce particular political economic formations, examined through what I have elsewhere termed computational capitalism (Berry 2025). The AI sprints methodology I propose here attempts to embed this into critical practice, developing reflexive critique in relation to the use of computational augmentation.

The time compression in sprint methodologies is designed to create specific epistemic affordances. As in data visualisation, certain patterns might become visible only through particular modes of temporal organisation. The intensive time-boxing of AI sprints similarly produces insights that emerge through rapid iteration rather than open-ended contemplation. However, this compression also risks a kind of analytical presentism, where the pressure to produce outputs within the available time privileges patterns more easily extractable by computational processing over those requiring sustained interpretive attention. However, by drawing on traditions established through FLOSS Manuals and the Book Sprint method, particularly through the work of Adam Hyde, AI sprints can mitigate these issues in order to developing rapid, collaborative knowledge production.

As detailed in *On Book Sprints* (Berry et al 2014), these intensive writing workshops emerged from the free software documentation movement, developing patterns for *extractive* knowledge capture and *generative* concept development. The Book Sprint method demonstrated that compressed timeframes, when properly facilitated, could produce high-quality outputs through intensive collaboration. AI sprints extend this process into computationally-supported research, replacing some human collaborators with LLM interlocutors whilst maintaining the emphasis on intermediate objects, iterative refinement, and critical reflexivity that characterises Book Sprint practice.

**Section 1: The Three Modes of Cognitive Augmentation**

My experimentation in (extended) critical code studies helped me to understand that cognitive augmentation is not a single approach but operates through three distinct modes which I now outline in more detail (see also Berry 2025b). Understanding these modes is helpful for designing effective AI sprint practices. These modes might be understood through what Suchman (2012) calls *configuration*, the combined shaping of human practice and machine processing through interaction. Each mode represents a different configuration of human-AI relations, with distinct distributions of agency and accountability. Here I develop this further by connecting it to broader questions of computational ontology and the politics of abstraction.

---

[6] The collaborative process I describe here is applied to text-based LLMs used for text and coding tasks. But mutatis mutandis they also apply to image generation systems like Stable Diffusion even though they use different technical architectures (e.g. iterative denoising processes rather than token prediction). The cognitive modes I identify, cognitive delegation, productive augmentation, and cognitive overhead operate in a slightly different register when the intermediate objects are images rather than code or structured data as visual outputs are more difficult to apply iterative revisions to and lack the version control that textual materials more readily accommodate.

## Cognitive Delegation

*Cognitive delegation* represents the primary issue to be avoided in AI-augmented research. This occurs when researchers offload interpretive judgement to the LLM, accepting its analytical framings uncritically. As I observed in my first attempt at software development, the ease of offloading problems to the LLM created a false confidence in using the AI. I had assumed the LLM could parse complex, semi-structured documents because previous prompts had succeeded with simpler tasks. The LLM, for its part, generated increasingly complex solutions without signalling that the approach was actually unworkable. In effect, the AI produced increasingly buggy, broken code, which it nevertheless cheerfully recommended to me to solve my problem.

This dynamic reveals how cognitive augmentation can easily become cognitive delegation so that rather than improving research project decisions, the LLM's positive responses encouraged me to persist with what turned out to be an unworkable strategy. In research contexts this becomes particularly hazardous as it wastes considerable amounts of time and takes the research down lines that will be fruitless. Where buggy code fails visibly, the analysis may nonetheless appear superficially plausible. The researcher must therefore take a stance of critical distance from the LLM and its *apparent capabilities*, recognising that its willingness to generate a solution or digital tool does not indicate the approach is sound.

For example, a researcher using an LLM to thematically code interview transcripts about precarious labour, might generate categories such as "job insecurity", "financial stress", "wellbeing" or "work-life balance" that appear helpful. Yet these themes may actually reflect the model's training on management consultancy documents and self-help literature rather than critical scholarship. A researcher who accepts these categories uncritically might therefore import a depoliticised framing that obscures structural questions about capitalism and power that a critical sociology would foreground. The codes appear plausible because they match commonsense understandings of work, yet this very plausibility conceals their theoretical inadequacy. As can be seen, cognitive delegation can enable a subtle theoretical drift towards frameworks already dominant in the LLM's training corpus without producing visible errors.

This process of abstraction operates, following Philip Agre (2014), through both "capture" in the acquisition of data and in the creation of ontologies that model the research domain (see Berry 2017). The danger in cognitive delegation lies not simply in offloading tasks to the LLM but in unconsciously accepting the epistemological and ontological presumptions embedded in its processing. This is similar to what Bowker and Star (2000) identify as the politics of classification, where seemingly neutral taxonomic decisions actually encode particular worldviews that then structure what becomes thinkable within a system. As I have argued, "procedures of abstraction make different knowledges comparable, calculable, and subject to re-engineering and reconstruction: they radically reshape the world in terms of the model that was originally abstracted" (Berry 2017: 104). Therefore, when researchers delegate analytical decisions to LLMs, they implicitly accept the ontological assumptions encoded in the model's training and architecture and these may subtly undermine humanistic and social scientific research paradigms.[7]

This danger is an example of what Weizenbaum (1976) identified as the "ELIZA effect", where humans readily attribute understanding to simple pattern-matching scripts. As my colleagues and I discovered through uncovering ELIZA's original source code, Weizenbaum's program operated through remarkably minimal linguistic transformations, yet users invested it with therapeutic insight and emotional intelligence (Berry and Marino 2024; see also Ciston et al 2026). The kind of "vibe coding" approach I adapt here for research purposes ("vibe research") represents a particularly interesting example of this phenomenon, as the tight iterative loops that make this method productive also create conditions where a positive assumption about AI competence can most easily take hold. Where the ELIZA effect described users misrecognising computational processing as understanding, similarly with LLMs, a "*competence effect*" emerges when the positive over-confidence of an LLM chatbot obscures or hides the absence of ability (Berry 2025).

The competence effect is potentially more problematic than the ELIZA effect because the evidence appears to support the researcher's anthropomorphism. The LLM produces analysis that reads plausibly, citations that look correct, and interpretations that seem insightful, creating a feedback loop that reinforces the users assumptions about the system's capabilities. Unlike ELIZA users who had to ignore the system's obvious repetitiveness and mistakes, researchers using AI sprints often receive constant positive reinforcement that the LLM "gets it". The competence effect thus represents a qualitatively different epistemological trap, one where generative ability becomes confused with real comprehension. I argue, therefore, that the competence effect operates differently from traditional notions of anthropomorphism in human-computer interaction because it emerges not from designers deliberately creating human-like interfaces (as with ELIZA's therapeutic approach) but from the LLM's training on vast human-generated text corpora. The system's outputs bear surface similarity to human reasoning not through programmed imitation but through statistical learning from human language patterns. This might be said to create anthropomorphism as an emergent property rather than designed feature, albeit shepherded using guard-rails and "system prompts." This might therefore make it more difficult for researchers and other users to maintain a critical distance from the software.

---

[7] This is linked to concerns that connect to Frankfurt School critiques of instrumental rationality more broadly (Berry 2014). As Horkheimer and Adorno argued, the subsumption of qualitative difference under quantification represents a key moment in the dialectic of enlightenment itself (Horkheimer and Adorno 2002). This is where reason's emancipatory promises become means for new forms of domination. LLM-mediated research risks accelerating this tendency, as the ontological commitments required for computational processing favour those forms of knowledge amenable to quantification, formalisation, and algorithmic manipulation (see Berry 2014, Berry 2023).

The competence effect might be understood as similar to what Noble (2018) describes as *technological redlining*. This is where computational systems appear to function neutrally whilst actually encoding particular assumptions and exclusions. Where Noble focuses on how algorithmic systems discriminate, the competence effect operates through *apparent inclusion*, creating the impression that the LLM genuinely understands research domains when it merely patterns responses based on training data. This makes the competence effect particularly problematic, as the researcher receives constant reinforcement that appears to validate their analytical approach whilst potentially reproducing unexamined assumptions embedded in the model.

**Productive Augmentation**

It is important therefore to develop a working method with AI that enables what I call *productive augmentation*. This is where researchers maintain strategic control over research questions, analytical frameworks, and interpretive claims whilst using LLMs to rapidly process large datasets, generate intermediate objects such as visualisations, summary tables, and extracted themes, and implement analytical procedures that would otherwise require extensive technical expertise or research assistance.

In my second attempt at software development, this mode emerged clearly. My realisation that the LLMs text extraction code was faulty, enabled me to direct the LLM towards a better solution. The LLM then augmented my implementation suggestions, rapidly generating the interface code I proposed but would have taken considerable labour to write manually. This, I think, usefully demonstrates the optimal human-LLM division of labour, where a human's capacity for creative thinking and seeing the whole combines with AI generation of implementation options.

For AI sprints, productive augmentation could mean that the researcher controls the research design, identifies appropriate analytical frameworks, determines what intermediate objects will make findings legible, and maintains interpretive control over the results. The LLM then handles the implementation through creation of digital tools or direct processing of texts, generating provisional visualisations, creating computer code, and producing summary information. By these means the augmentation remains supplementary rather than substitutive, with each actor, human or AI, contributing capabilities the other lacks.

As Hayles (2012) observed, we think through, with, and alongside media. Productive augmentation brings this idea to the fore, allowing the LLM to become a medium for thought that extends human analytical capacity without displacing human judgement. However, maintaining this mode requires constant attention not to drift into cognitive delegation.

One way of doing this is to create initial "prompt protocols" that help structure the human-LLM interaction. For example, (1) always specify the analytical framework before requesting processing ("using Bourdieu's field theory, extract..."), (2) request multiple alternative approaches to prevent premature analytical closure ("generate three different thematic schemes for..."), (3) demand transparency about processing decisions ("explain what criteria you used to..."), and (4) treat LLM outputs as *provisional materials* requiring critical evaluation. These protocols help to foreground the division of labour between human theoretical work and computational processing, preventing the gradual drift towards cognitive delegation that tight iterative loops might otherwise encourage.

**Cognitive Overhead**

The third mode I want to discuss I call *cognitive overhead* and represents the scaling limits of AI augmentation. As I soon discovered in this software development project, beyond certain complexity levels the mental resources required to manage LLM context may exceed resources saved by not writing code directly. For example, the context window for my conversation with Gemini became increasingly unwieldy as the software grew more complex. The LLM began to get confused about the version we were working on, attempted to reimpose previously rejected features, or rolled back to earlier versions. Careful prompting was required to remind it which version was current, phrases like "fix only this, update the version number", "add the 2.4 comment to this file", etc. Indeed, I had to suggest to the AI the use of version numbering as it couldn't keep track of different iterations that it was generating. The cognitive resources freed by not writing implementation code were partially consumed by managing the LLM's context and preventing reversions in the project.

This is distinct from both delegation and productive augmentation because rather than augmenting capabilities, the LLM now requires active human management and oversight. This included tracking version numbers, constraining scope with phrases like "do only these changes", and preventing unsolicited modifications. The creative freedoms opened up by not writing implementation code are gradually consumed by project managing the LLM's work. This shows that cognitive augmentation is not simply additive but involves real trade-offs where automating research and tools must be set against the cost of AI management.

For an AI sprint method, cognitive overhead suggests scaling limits beyond which the LLM will simply be unable to cope and require human-support to continue to work effectively. The approximately 1000-line program I developed reached these limits surprisingly quickly. For complex research projects involving multiple analytical threads, managing LLM context across extended timeframes may become more exhausting for the researcher than the gains from AI augmentation. This suggests AI sprints work best for time-boxed, intensive research tasks of perhaps 1-5 days duration, where the entire project remains manageable within an LLM context window. Extended scholarly projects spanning weeks or months would require either breaking the work into discrete sprint episodes, each producing intermediate objects that feed subsequent sprints, or adopting hybrid approaches that combine AI-augmented analysis with traditional inquiry. But we should also remember that the LLMs are developing quickly in their capacities, and this kind of meta-project knowledge will surely be a key objective for them to be able to manage in some way. At present we see attempts through Retrieval-Augmented Generation (RAG) and Model Context Protocol (MPC) technologies, together with interface tools like Cursor, to help manage these issues, but they remain a very real problem as the project grows beyond a certain threshold which is effectively an AI context collapse.

## Section 2: Key Principles

Before outlining the key principles, it is important to acknowledge the material conditions of possibility for an AI sprint practice. Whilst I have tried to show how AI augmentation can reduce dependence on technical expertise, this apparent democratisation may conceal new forms of infrastructural dependence. Access to frontier LLMs typically requires institutional subscriptions, API credits, or acceptance of platform terms of service that may restrict research freedom. The apparent reduction in human collaboration costs is thereby offset by computational costs that may be unevenly distributed across institutions and geographical regions. Moreover, the proprietary nature of most frontier LLMs means researchers depend on corporate platforms whose priorities often are very different from scholarly values. A critical AI sprint methodology must therefore be reflexive about these political economic issues, considering how platform dependencies shape what research becomes possible and for whom.

These infrastructural dependencies connect to broader questions about computation and capitalism, where value extraction increasingly depends on controlling computational infrastructure or platforms. AI sprints operate within a political economy that shapes not just what research costs but also what forms of inquiry become viable. A critical methodology must therefore acknowledge how platform capitalism structures the horizon of possible research, making certain questions easier to pursue than others based not on intellectual merit but on how commercial incentives have shaped the way algorithms are designed and implemented in machine learning systems.

AI sprints also raise novel ethical questions about research subjects and data. When LLMs are used to process interview transcripts, archival materials, or ethnographic field notes, they potentially expose research participants to forms of analysis usually not anticipated in ethical consent. For example, the circulation of prompts and intermediate objects as part of methodological transparency may accidentally reveal identifying information about participants. A critical AI sprint should therefore reflect on the ethical issues that it might raise. These might include, (1) whether existing consent approval or procedures adequately cover LLM processing of qualitative data, (2) how to anonymise intermediate objects whilst preserving their utility, (3) whether participants should be informed that AI systems will process their contributions, and (4) how to handle cases where LLM outputs contain hallucinated but plausible-sounding claims about research participants.[8]

I now want to briefly outline what I think are the key principles that need to be kept in mind when developing an AI sprint. Firstly, it is crucial that some sense of *architectural control* is maintained by the researcher to ensure that strategic authority over research questions, analytical frameworks, and interpretive claims is ensured. The idea is that LLMs are used to create or generate research elements but do not determine the research direction. As I argue elsewhere, this requires treating the LLM as "a sophisticated pattern-completion engine rather than a genuine collaborator with understanding" (Berry 2025b). Secondly, it is important to foreground the value of producing *intermediate objects*, such as visualisations, summary tables, coded excerpts, thematic mappings, extracted JSON files, etc. that make LLM processing legible and contestable by the researcher. I think this might be one of the most important insights I have learnt from undertaking an AI Sprint. Researchers often assume LLMs can jump this stage and produce comprehensive research outputs – but in fact this is far from the truth due to limitations in the AIs we currently have. Creating these intermediate objects also prevent cognitive delegation by making algorithmic analysis visible and subject to critical analysis. As Latour (1999) demonstrates through his concept of "circulating references", scientific facts emerge through transformations that progressively abstract from material phenomena to inscriptions. AI sprints can make this process very explicit through a process of raw data becoming intermediate objects (via the AI) then becoming analytical claims, with each transformation open to examination by the researcher.

---

[8] The issue of hallucinations by LLMs is important because when processing research data, fabricated patterns or quotes may appear alongside real analysis, creating risks of misrepresentation that usually isn't a problem in qualitative research using source materials. Extra care will therefore need to be taken to "sanity-check" intermediate outputs and final results. Additionally, when researchers process sensitive qualitative data through commercial AI platforms, they potentially expose participants to corporate data harvesting practices whose full scope remains opaque. This suggests AI sprint methods may be inappropriate for particularly sensitive research, such as studies of marginalised communities, political dissent, or criminalised practices, where even the possibility of corporate data harvesting poses unacceptable risk. Together, these raise important ethical issues that may need to be carefully reviewed before a project is undertaken.

We can understand intermediate objects in AI sprints not merely as transitional representations but as "materialised abstractions" that actively co-structure the research process. These objects emerge through what I have described as "a grammatisation process that discretises and re-orders" research materials (Berry 2016: 108), transforming complex qualitative materials into computationally usable forms and enabling new possibilities for analysis. The concern lies in what I call the technical "derangement of knowledge" where computational processing introduces distance from the source materials so that human intelligibility and interpretability is undermined (Berry 2016: 109). The more processing required to make materials computationally tractable the further removed they become from their original contexts, yet the danger is that they appear as objective facts rather than constructed representations. For example, when an LLM is used to process interview transcripts to generate a thematic summary table, the grammatisation operates through tokenisation, embedding, and pattern extraction that necessarily distances the intermediate object from the semantic thickness of the original conversation. The table generation makes certain patterns computationally legible whilst rendering other dimensions of meaning opaque or absent. The researcher has to remain critically aware not simply of what the table shows but what processes of abstraction produced it. This is to take an interpretative stance that I argue is key within critical digital humanities (Berry 2023).

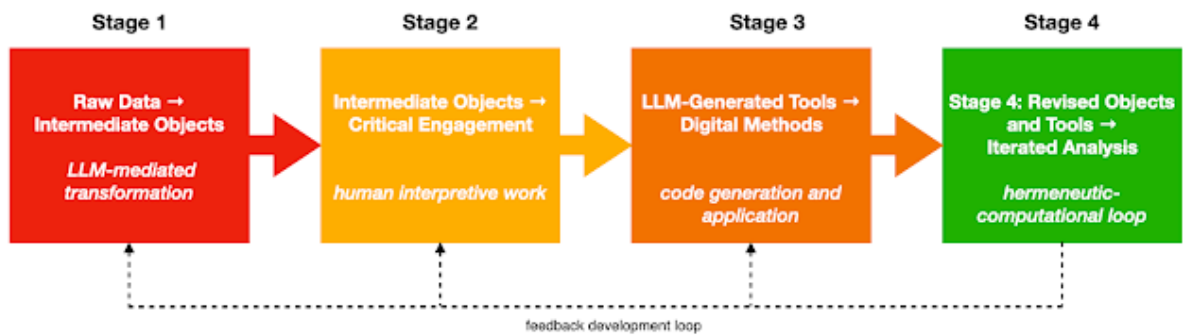

*Figure 3: The hermeneutic-computational loop between human understanding and computational generation[9]*

---

[9] Figure 3 represents the iterative cycle central to the AI sprint practice, showing the dynamic relationship between human interpretive judgement and generative AI. The loop takes place through four key moments: (1) the researcher formulates analytical questions and prompts based on domain expertise and theoretical frameworks, (2) the LLM processes materials and generates intermediate objects such as visualisations, summary tables, or coded excerpts, (3) the researcher critically examines these outputs, identifying patterns, anomalies, absences, and questions requiring further investigation, (4) insights from this examination inform revised prompts, alternative analytical approaches, or reorganisation of data structures, reinitiating the cycle. This research loop encourages productive augmentation rather than cognitive delegation by foregrounding human strategic overview whilst utilizing LLMs for normalisation, processing and generation of data, texts and images. It is interesting to note that meaning emerges from the ongoing dialectical engagement, with each iteration potentially refining both the computational processing and human understanding.

Thirdly, it is important to develop a process of *iterative revision* where AI sprints use tight feedback loops between human interpretation and AI processing (see Figure 3). The researcher needs to pay careful attention to the intermediate objects generated, and identify patterns, anomalies, and questions requiring further investigation, then revise the prompts being used. This can involve introducing different analytical approaches, alternative visualisations, targeted probes into the data, or reorganising the data created (e.g. differently ordered JSON files). This enables what I term a *hermeneutic-computational loop* between human understanding and computational generation (Berry 2025b).

The next key principle is the importance of *methodological transparency* which is made possible by the textual outputs of LLMs which enables the documenting of prompts, LLM interactions, and key decision points or failures. This enables the possibilities of reproducibility, critique, and collective methodological learning by sharing and reflecting on these materials. As Marino (2020) has argued for in relation to the method of critical code studies, we must attend not just to what systems do but what they mean, examining how analytical choices shape possible interpretations. The last key principle is the need for continual *reflexive critique* throughout an AI sprint. This requires constant care to guard against this notion of the competence effect. It is important that researchers direct critical attention to the seductive effect of speed and scale, which may come at a cost of nuance, scholarly rigour and attention to context in the research. Indeed, an LLM's capabilities are usually unevenly distributed across different analytical tasks, which makes it more suited to some than others. In addition, an AI's training data may shape what becomes analytically visible or invisible in unexpected ways due to biases, absences and overfitting of the data.

In the interests of transparency when undertaking AI sprints, it is useful to document failures and abandoned approaches. For example, when an LLM is unable to handle a particular analytical task or when prompts fail to generate useful intermediate objects, they can be very useful for helping to explain why certain methods were used or results obtained. Failures, such as these, can reveal the epistemological limits of computational processing more clearly than successes, showing which aspects of humanistic inquiry required additional interpretive work. An AI sprint should therefore record not just results but also its breakdowns and failures.

**Conclusions**

AI sprints are a specific method for AI-augmented research, but I think their wider contribution lies in what I am calling "critical augmentation" as a research ethos.[10] I've tried to show that computational

---

[10] The AI sprint approach might also be applied reflexively to the study of artificial intelligence systems themselves, where AI augmentation becomes both method and object of inquiry. For instance, researchers might use LLMs to process large

systems can extend human analytical capacities whilst insisting that such augmentation requires constant reflexive vigilance against instrumental rationality's encroachment on the research process. As I discovered through developing an example of educational software via "vibe coding", collaboration with LLMs produces useful insights whilst requiring constant care and project management combined with critical interpretability (Berry 2025b). LLMs can process texts at scales impossible for individual scholars, generate visualisations revealing unexpected patterns, create interfaces without requiring technical expertise, and so produce bespoke computational tools from natural language prompts.[11] Yet they cannot determine whether these processes are intellectually sound, contextually appropriate, or interpretively useful. That judgement remains with the human researcher.

The critical augmentation I am arguing for here is part of a necessary resistance to computational thinking's hegemonic claims. AI sprints show that LLMs can augment humanistic inquiry without displacing the interpretive practices central to critical scholarship. However, this requires constant vigilance against computational solutionism, that is, the assumption that every research problem has a computational answer. As Winner (1980) argued regarding technology more broadly, the question is not whether computational augmentation is possible but whether it serves humanistic values and intellectual autonomy. AI sprints work when they extend rather than replace human judgement. This is possible when they make visible rather than obscure their analytical operations rather than simply accelerating existing practices and when they enable research questions we might not otherwise pursue.

We might note that the wider significance of AI sprints, perhaps, extends beyond their immediate use as a method. As LLMs become increasingly capable, the questions they raise about cognitive delegation, interpretive authority, and intellectual autonomy will only become more pressing. We are nearing a future where LLMs can not only process data in a number of remarkably useful ways, but

---

corpora of AI safety literature, technical documentation, or platform terms of service (ToS), generating intermediate objects that help to reveal patterns in how AI systems are governed, marketed, or constrained. Similarly, AI sprints could analyse datasets of LLM outputs to study the competence effect empirically, using computational methods to identify where generative output regresses away from accuracy or coherence. Researchers might also use AI sprints to examine training data biases, prompt engineering practices, or the circulation of AI-generated content across platforms, with the LLM serving as both analytical tool and research object.

[11] Particularly promising is the possibility to use AI sprints to map the probabilistic outputs of AI outputs through sophisticated version variation type approaches. This might operate in a number of ways. For example, researchers could use one LLM (the "research AI") to systematically control and question another LLM (the "subject AI"), running identical prompts iteratively to capture the stochastic variation in outputs of synthetic media or AI slop (see Berry 2025). This would make visible the deterministically non-deterministic character of these systems, revealing how temperature settings, sampling methods, and hidden parameters shape the distribution of possible outputs. The research AI could also generate intermediate objects such as variation matrices, semantic clustering tables, or divergence/difference metrics that document the boundaries of output stability. Another approach might be to use an AI sprint to systematically compare different versions of particular frontier AIs (e.g. GPT-4 versus GPT-4.1, or Claude Opus versus Claude Sonnet), documenting how architectural changes, fine-tuning, or safety guardrails alter outputs. Through a comparative analysis one might reveal how commercial pressures, regulatory concerns, or user feedback reshape AI behaviour across version iterations, providing empirical grounding for claims about AI governance and development.

also potentially generate entire research papers, propose theoretical frameworks, and even undertake a form of artificial "peer review". In this situation, the principles developed through AI sprint methods, particularly the insistence on human strategic overview and the production of legible intermediate objects, become crucial for maintaining the distinction between *AI-augmented scholarship* and *AI-generated simulacra of scholarship*. AI sprints can therefore be seen as an early attempt to establish methods and practices that ensure human judgement remains central to knowledge production, even as computational systems become ever more sophisticated in their imitative capabilities. The AI method's value lies not in efficiency but in opening new possibilities for critical work whilst maintaining the reflexivity essential to humanistic inquiry.